\documentclass[graybox]{svmult}
\usepackage{newtxtext}
\usepackage{newtxmath}
\usepackage{helvet}         
\usepackage{courier}        
\usepackage{type1cm}        
\usepackage{makeidx}         
\usepackage{graphicx}        
\usepackage{multicol}        
\usepackage[bottom]{footmisc}
\usepackage{tikz} 

\usepackage[dvipsnames]{xcolor}
\usepackage{comment}

\usepackage{url}

\usepackage{mathtools}
\usepackage{bm}
\usepackage{microtype}
\usepackage{hyperref}

\begin{document}

\title{Numerical Solution of the Bardeen--Cooper--Schrieffer Equation for Unconventional Superconductors}
\titlerunning{Bardeen--Cooper--Schrieffer Equation} 
\author{Andreas A. Buchheit, Torsten Ke{\ss}ler and Sergej Rjasanow}
\authorrunning{A. A. Buchheit et al.}
\institute{A. Buchheit \at Saarland University, 66123 Saarbr\"ucken, Germany \at \email{andreas.buchheit@uni-saarland.de} 
\and T. Ke{\ss}ler \at Eindhoven University of Technology, 5612 Eindhoven, Netherlands
\and S. Rjasanow \at Saarland University, 66123 Saarbr\"ucken, Germany}

\maketitle

\abstract{
In this work, we consider the analytical properties and the efficient numerical solution of the Bardeen–Cooper–Schrieffer equation for unconventional superconductivity incorporating long-range power-law electron–electron interactions  within a tight-binding model on a $d$-dimensional lattice. It is a nonlinear convolution equation for the complex matrix-valued superconducting gap under symmetry constraints imposed by the fermionic anticommutation rules. The long-range interaction enters in momentum space in the form of the now efficiently computable Epstein zeta function, which exhibits a power-law singularity at zero momentum. This needs to be accounted when evaluating the convolution. After a brief overview of some of the equation's analytical properties, we discuss its efficient numerical solution using a Galerkin method with B-splines. We present numerical results for a nodal superconductor on a two-dimensional square lattice.
} 


\section{Introduction}
\label{Sec: Introduction}

Superconductivity is a quantum phenomenon in which electric resistance vanishes below a certain critical temperature. It has a wide range of applications, ranging from energy transport to quantum computation. After being discovered in 1911 by Onnes, the first microscopic description of superconductivity was provided by Bardeen, Cooper, and Schrieffer in 1957 \cite{BCS}. They showed that an attractive interaction mediated by phonons between electrons could lead to the formation of Cooper pairs. The central mathematical quantity of interest here is the superconducting gap, which contains the information about the energy required to break the Cooper pair apart. This quantity can be obtained as the solution of the nonlinear Bardeen-Cooper-Schrieffer (BCS) gap equation, where a non-zero stable solution indicates the presence of superconductivity.   Since the discovery of high-temperature superconductors in 1986 and other unconventional superconductors, there has been a need for a better understanding of their microscopic origins beyond phonon-mediated interactions \cite{sigrist1991phenomenological,stewart2017unconventional}. Recently, some of the authors have shown that long-ranged electron-electron interactions can lead to topological superconducting states, with potential applications in quantum computing \cite{buchheit2023exact,haink2025non}.

Compared to the volume of published physical papers on the topic, works on the mathematical and numerical aspects of superconductivity are scarce. The analytical properties of the BCS functional have been discussed by Hainzl and Seiringer in \cite{hainzl2016bardeen}. The effect on the critical temperature due to the presence of boundaries was analysed within a continuum approximation in \cite{hainzl2023boundary}. The numerical solution of inhomogeneous BCS superconductors  in real space using a Chebyshev basis was discussed in \cite{covaci2010efficient}. A recent energy functional-based approach that is also applicable to repulsive interactions has been recently provided in \cite{fanfarillo2025superconductivity}.

Let $\Lambda = A \mathbb Z^d$ be a $d$-dimensional lattice with $d=1,2,3$ and $A\in \mathbb R^{d\times d}$ regular. For $\mathbb{T}^d=[0,1)^d$ the $d-$dimensional torus, the unit cell of $\Lambda$ is defined as $E=A\mathbb T^d$ with volume $V_\Lambda=|\det A|$. We further define the reciprocal lattice $\Lambda^\ast$ with reciprocal unit cell  $E^\ast=A^{-T} \mathbb T^d$. We finally consider the unknown $F\in L_2(E^\ast,\mathbb C^{k\times k})$,  with $k=1,2$. Here, $F$ denotes the superconducting gap matrix as a function of the relative momentum of two electrons. Further, $k=1$ corresponds to the scalar case where only singlet superconductivity is considered, whereas $k=2$ includes the spin degree of freedom and thus both the cases of singlet and triplet electron-electron pairing.

We consider the nonlinear BCS convolution equation for superconductors with attractive long-range interactions \cite{BCS,buchheit2023exact}
\begin{equation}
	\label{Eqn: Main Equation F}
	F(\boldsymbol{x})=
	\int\limits_{E^\ast}K(\boldsymbol{x}-\boldsymbol{y})G[F](\boldsymbol{y})
	\mathrm{d}\boldsymbol{y}
\end{equation}
for the function $F$, where in the case of triplet pairing ($k=2$) the additional symmetry constraint 
\begin{equation}
\label{eq:symmetry}
F^\top(-\bm x)=-F(\bm x)
\end{equation}
needs to hold.
The kernel $K$ is of the form
\begin{equation}
	\label{Eqn: Kernel}
	K(\boldsymbol{x}-\boldsymbol{y})=C_1+C_2\,
	Z_{\Lambda,\nu}(\bm x-\bm y)
\end{equation}
where $C_1\ge 0$ is the so-called on-site interaction and $C_2\geq 0$ the prefactor of the long-range interaction per reciprocal lattice cell volume. Here, we adopt dimensionless units, writing energies in units of $\hbar \tau$ with $\tau$ the electron hopping amplitude and $\hbar$ the reduced Planck constant and positions in units of the smallest distance between lattice sites $a_0$. In momentum space, the power-law interaction with exponent $\nu\in \mathbb C$ on the lattice $\Lambda$ is encoded in the Epstein zeta function $Z_{\Lambda,\nu}$
\cite{epstein1903theorieI,epstein1903theorieII}, with
\begin{equation}
	\label{Eqn: Epstein zeta function}
	Z_{\Lambda,\nu}(\bm y)
	=\,\sideset{}{'}\sum_{\boldsymbol{z}\in\Lambda}
	\frac{e^{-\imath\,2\pi\,\boldsymbol{y}\cdot\boldsymbol{z}}}
	{|\boldsymbol{z}|^\nu},\,\quad \bm y\in E^\ast,\quad \operatorname{Re}(\nu)>d,
\end{equation}
where the primed sum excludes the case $\bm z=\bm 0$. The function can be meromorphically continued to $\nu\in \mathbb C$. Its analytical properties and the efficient numerical evaluation of the Epstein zeta function have recently been described in \cite{buchheit2024epstein}, together with an implementation in the high-performance C-library EpsteinLib, which is publicly available on \href{https://github.com/epsteinlib/epsteinlib}{GitHub}.

Finally, the non-linear mapping $G$ is defined as follows
\begin{equation}
	\label{Eqn: Direct F->G}
	G[F](\bm y)=F(\bm y)\,\big(\xi^2(\bm y)I+F^*(\bm y)F(\bm y)\big)^{-1/2}\,,
\end{equation}
with $I$ denoting the $k\times k$ identity matrix and with $F^\ast$ denoting the adjoint matrix. Further, the scalar function $\xi\,:\,E^\ast\rightarrow \mathbb{R}$, called the dispersion relation,
describes hopping of electrons between lattice sites. In case of nearest-neighbor hopping, it is given
by
\begin{equation*}
	\label{Eqn: Non-linear mapping}
	\xi(\bm y)=-\sum_{j=1}^{d}\cos(2\pi (A \bm y)_j).
\end{equation*}
Note that the inverse square root in \eqref{Eqn: Direct F->G} is well defined for all $\bm y$ where $\xi(\bm y)\neq 0$, since the argument is positive definite. On the $(d-1)$-dimensional Fermi surface set $\mathcal F=\{\bm y: \xi(\bm y)= 0\}$, the function $G[F](\bm y)$ can, however, develop non-smooth behavior when $F(\bm y)$ loses rank. This is referred to as nodal superconductivity. However, since the Fermi surface has zero Lebesgue measure, the convolution is well-defined as the integrand is defined almost everywhere.

\section{Theoretical study}
For ease of notation, we will focus on the case of a square lattice, $A=I$, in what follows. The results, however, generalise under minor modifications to arbitrary lattices $\Lambda$. We begin our study with the simplest scalar one-dimensional ($d=k=1$) an a constant kernel, i.e. $C_1>0,\, C_2=0$. Thus, we get
\begin{equation*}
\label{Eqn: 1D-case}
f(x)=C_1\rho[f](x)=C_1\int\limits_0^1\dfrac{f(y)}{\sqrt{\cos^2(2\pi\,y)+|f(y)|^2}}\mathrm{d}y.
\end{equation*}
The solution $f$ is then a complex constant. Its exponential form $f=s\,e^{\imath\,\phi}, s\ge 0, \phi\in\mathbb{R}$ leads to a real-valued fix point equation
\begin{equation}
	\label{Eqn: 1D-case real}
	s= C_1\int\limits_0^1\dfrac{s}{\sqrt{\cos^2(2\pi\,y)+s^2}}\mathrm{d}y=C_1\,\varphi(s)\,.
\end{equation}
The function $\varphi$ allows an analytical representation
$$
\varphi(s)=\int\limits_0^1
\frac{s}{\sqrt{\cos^2(2\pi\,y)+s^2}}\,
\mathrm{d}y=\frac{1}{\pi}
\left(
K\Big(-\frac{1}{s^2}\Big)+\frac{s}{\sqrt{1+s^2}}K\Big(\frac{1}{1+s^2}\Big)
\right)\,,
$$
where $K$ denotes the complete elliptic integral of the first kind defined for the variable $-\infty < m <1$ as
$$
K(m)=\int\limits_0^{\pi/2}\frac{\mathrm{d}y}{\sqrt{1-m\sin^2(y)}}\,.
$$
Thus, the function $\varphi$ follows the non-trivial asymptotic for $s\rightarrow 0^+$
$$
\varphi(s)=-\frac{2}{\pi}\,s\,\ln s+
\frac{4\ln 2}{\pi}\,s+\mathcal{O}(s^3\ln s)\,.
$$
Its further properties are
\begin{enumerate}
	\item[1.] $0\le\varphi(s)<1\,,\quad \varphi(0)=0\,,\quad \lim\limits_{s\rightarrow \infty}\,\varphi(s)=1$\,,
	\item[2.] $\varphi'(s)>0\quad $for all$\quad s>0\quad $with$\quad \lim\limits_{s \to 0} \varphi'(s) = \infty$.
\end{enumerate}

\begin{figure}
	\begin{center}
		\includegraphics[width=8.5cm]{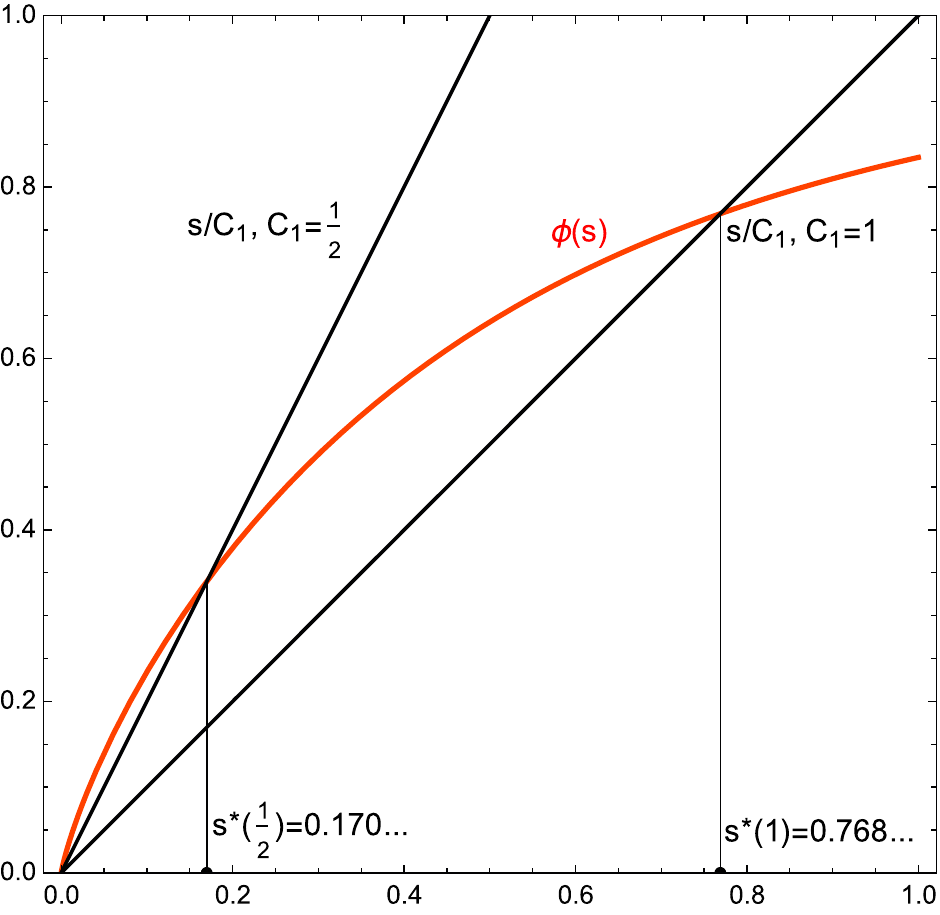}
	\end{center}
	\caption{Equation $s/C_1=\varphi(s),\ C_1=1,\ C_1=1/2$ for $d=1$.}
    \label{Fig:s-solution}
\end{figure}

Thus, there are exactly two intersections of the straight line $s/C_1$ and of the function $\varphi(s)$ at $s^*=0$ and at $0<s^*<C_1$ as it is illustrated in Fig.~\ref{Fig:s-solution}. The nontrivial solution $s^*$ with the asymptotic
$$
 s^*(C_1)\approx 4\,e^{-\pi/(2C_1)}\,,\quad C_1\rightarrow 0\quad \text{and}\quad
  s^*(C_1)\approx C_1\,,\quad C_1\rightarrow \infty
$$
leads to a variety of solutions of the equation (\ref{Eqn: 1D-case}) of the form $f=s^*e^{\imath\,\phi}, \phi\in\mathbb{R}$.
In the multidimensional ($d>1$) scalar case ($k=1$), the situation is similar. The function 
$$
 \varphi(s)=\int\limits_{\mathbb{T}^d}\frac{s}{\sqrt{\xi^2(\bm y)+s^2}}\mathrm{d}\bm y
$$
has the same properties as in the one-dimensional case. Therefore, there are two solutions: the trivial solution $f=0$, and non-trivial solution $f=s^*e^{\imath\,\phi}$.
The matrix-valued case $k=2$ for constant $F$ is equivalent to $k=1$, as we show in the following. The equation for the constant matrix $F\in\mathbb{C}^{k\times k},\, k>1$ now reads as follows
$$
 F=C_1\int\limits_{\mathbb{T}^d}F\,\big(\xi^2(\bm y)I+F^*F\big)^{-1/2}\mathrm{d}\bm y\,.
$$
By the use of the singular value decomposition $F=U\Sigma_FV^*$, we obtain
$$
U\,\operatorname{diag}\Big(\sigma_j-C_1 \int\limits_{\mathbb{T}^d}
\frac{\sigma_j}{\sqrt{\xi^2(\boldsymbol{y})+\sigma_j^2}}\,
\mathrm{d}\boldsymbol{y}\Big)_{j=1}^k\,V^*=0\,,
$$
and, since the matrices $U$ and $V^*$ are regular, to a set of independent algebraic equations
$$
\sigma_j-C_1 \int\limits_{\mathbb{T}^d}
\frac{\sigma_j}{\sqrt{\xi^2(\boldsymbol{y})+\sigma_j^2}}\,
\mathrm{d}\boldsymbol{y}=0\,,\quad j=1,\dots,k\,
$$
having the solutions $\sigma_j=0$ or $\sigma_j=s^*$ for all $j$. Therefore,  for $k=2$, there are four possibilities for the matrix $\Sigma_F$
$$
 \Sigma_F=\begin{pmatrix} 0 & 0 \\ 0 & 0 \end{pmatrix},\quad
 \Sigma_F=\begin{pmatrix} s^* & 0 \\ 0 & 0 \end{pmatrix},\quad
 \Sigma_F=\begin{pmatrix} 0 & 0 \\ 0 & s^* \end{pmatrix},\quad
 \Sigma_F=\begin{pmatrix} s^* & 0 \\ 0 & s^* \end{pmatrix}\,.
$$
Under inclusion of the symmetry condition from \eqref{eq:symmetry}, these solutions reduce to $F(\bm x) =0$  and \[F(\bm x)=\begin{pmatrix} 0 & s^\ast \\ -s^\ast & 0 \end{pmatrix}\] up to multiplication with an arbitrary complex phase factor.

The second part of the kernel \eqref{Eqn: Kernel} corresponds to a one-periodic
pseudo-differential convolution operator 
\begin{equation}
	\label{Eqn: Psi-do operator}
	\mathcal{B}[G](\bm x)= \int\limits_{\mathbb{T}^d}Z_{\mathbb{Z}^d,\nu}
	({\bm x - \bm y})
	G(\bm y)\mathrm{d}\bm y\,.
\end{equation}
The theory of these operators is well known (see f.e. \cite{Vainikko}) and is based on the Fourier series and one-periodic Sobolev spaces on $\mathbb{T}^d$.
Consider the orthonormal basis
$$
\left\{v_{\boldsymbol{n}}(\boldsymbol{x})\right\}_{\boldsymbol{n}\in\mathbb{Z}^d}=
\left\{e^{\imath\,2\pi\boldsymbol{n}\cdot\boldsymbol{x}}\right\}_{\boldsymbol{n}\in\mathbb{Z}^d}\,.
$$
The Fourier series of the scalar-valued function $f$ belonging to $L_2(\mathbb{T}^d)$ is
\begin{equation}
	\label{Eqn: Fourier series}
	f(\boldsymbol{x})=\sum_{\boldsymbol{n}\in\mathbb{Z}^d}
	f_{\boldsymbol{n}}\,v_{\boldsymbol{n}}(\boldsymbol{x})\,,\quad
	f_{\boldsymbol{n}}=\langle f,v_{\boldsymbol{n}}\rangle_{L_2}=\int\limits_{\mathbb{T}^d}f(\boldsymbol{x})\,
	e^{-\imath\,2\pi\boldsymbol{n}\cdot\boldsymbol{x}}\mathrm{d}\boldsymbol{x}\,.
\end{equation}
By analogy, for any linear mapping $\psi\,:\,C^\infty(\mathbb{T}^d)\rightarrow \mathbb{C}$, we define
its Fourier coefficients as $\psi_{\boldsymbol{n}}=\psi(v_{-\boldsymbol{n}})\,,\ \boldsymbol{n}\in\mathbb{Z}^d\,.$ 

The following abbreviation will often be used: $\langle\boldsymbol{n}\rangle=(1+|\boldsymbol{n}|^2)^{1/2}\,,\ \boldsymbol{n}\in\mathbb{Z}^d\,.$
The Sobolev space ${H}^s={H}^s(\mathbb{T}^d)\,,\ s\geq 0$, is defined as the completion of ${C}^\infty(\mathbb{T}^d)$ with respect to the norm
\begin{eqnarray}
	\label{Eqn: Sobolev norm}
	\|f\|_{{H}^s}=
	\left(\sum_{\boldsymbol{n}\in\mathbb{Z}^d}\langle\boldsymbol{n}\rangle^{2s}|f_{\boldsymbol{n}}|^2\,\right)^{1/2}\,.
\end{eqnarray}
With the scalar product,
$$
	\langle f,g \rangle_{{H}^s}=
	\sum_{\boldsymbol{n}\in\mathbb{Z}^d}
	\langle\boldsymbol{n}\rangle^{2s}f_{\boldsymbol{n}}\,\overline{g_{\boldsymbol{n}}}\,,
$$
the Sobolev space ${H}^s$ is a Hilbert space. For $s<0$, the Sobolev space ${H}^s$ 
is the space of distributions over ${H}^{-s}$, i.e. the dual space, equipped with the norm
$$
	\|\psi\|_{{H}^s}=\sup_{f\in{H}^{-s}}\dfrac{|\psi(f)|}{\ \|f\|_{{H}^{-s}}}\,,
$$
where $\psi(f)$ can be computed by the use of the Fourier coefficients as
$$
\psi(f)=\sum_{\boldsymbol{n}\in\mathbb{Z}^d}f_{\boldsymbol{n}}\,\psi_{-\boldsymbol{n}}\,.
$$
Every linear operator $\mathcal{A}\,:\,{H}^s\rightarrow{H}^r$  for $s,r\in\mathbb{R}$
can be formally applied to the Fourier series of the function $f\in{H}^s$,
$$
\mathcal{A}[f](\boldsymbol{y})=\sum_{\boldsymbol{n}\in\mathbb{Z}^d}
f_{\boldsymbol{n}}\,\mathcal{A}[v_{\boldsymbol{n}}](\boldsymbol{y})=
\sum_{\boldsymbol{n}\in\mathbb{Z}^d}
f_{\boldsymbol{n}}\,\sigma_{\mathcal{A}}(\boldsymbol{y},\boldsymbol{n})\,v_n(\boldsymbol{y})\,,
$$
where $\sigma_{\mathcal{A}}\,:\,\mathbb{T}^d\times\mathbb{Z}^d\rightarrow\mathbb{C}$, defined via
$$
	\sigma_{\mathcal{A}}(\boldsymbol{y},\boldsymbol{n})=
	\overline{v_{\boldsymbol{n}}(\boldsymbol{y})}\mathcal{A}[v_{\boldsymbol{n}}](\boldsymbol{y})
$$
denotes the symbol of the operator $\mathcal{A}$. The usual assumptions on the symbol $ \sigma_{\mathcal{A}}$ are
\begin{enumerate}
	\item[1.] $\sigma_{\mathcal{A}}(\cdot,\boldsymbol{n})\in {C}^\infty(\mathbb{T}^d)\,,\quad 
	\forall\boldsymbol{n}\in\mathbb{Z}^d$\,,\vspace{1ex}
	\item[2.] $\big|\partial_{\boldsymbol{y}}^{\alpha}\Delta_{\boldsymbol{n}}^{\beta}
	\sigma_{\mathcal{A}}(\boldsymbol{y},\boldsymbol{n})\big|\leq C_{\alpha,\beta}\,
	\langle\boldsymbol{n}\rangle^{\gamma-|\beta|}$.
\end{enumerate}
Here, $\gamma\in\mathbb{R}$ is called the order of the operator $\mathcal{A}$, $\alpha,\beta\in\mathbb{N}_0^d$ are two multi-indices, $C_{\alpha,\beta}>0$ a family of positive constants, $\partial_{\boldsymbol{y}}^{\alpha}$ the usual partial derivative of the order $|\alpha|$, and
$\Delta_{\boldsymbol{n}}^{\beta}$ is a finite difference operator of order $|\beta|$.

Let $\mathcal{A}$ be an integral operator over $\mathbb{T}^d$ with
$$
	\mathcal{A}[f](\boldsymbol{x})=
	\int\limits_{\mathbb{T}^d}K(\boldsymbol{x},\boldsymbol{y})
	f(\boldsymbol{y})\mathrm{d}\boldsymbol{y}\,.
$$
Then, its symbol $\sigma_{\mathcal{A}}$ is given by
$$
	\sigma_{\mathcal{A}}(\boldsymbol{x},\boldsymbol{n})=
	e^{-\imath\,2\pi\boldsymbol{n}\cdot\boldsymbol{x}}
	\int\limits_{\mathbb{T}^d}K(\boldsymbol{x},\boldsymbol{y})
	e^{\imath\,2\pi\boldsymbol{n}\cdot\boldsymbol{y}}\mathrm{d}\boldsymbol{y}\,.
$$
In the special case of a convolution integral operator $\mathcal{A}$ with
$$
	K(\boldsymbol{x},\boldsymbol{y})=K(\boldsymbol{x}-\boldsymbol{y})\,,
	$$
	its symbol $\sigma_{\mathcal{A}}$ simplifies to
	$$
	\sigma_{\mathcal{A}}(\boldsymbol{x},\boldsymbol{n})= \sigma_{\mathcal{A}}(\boldsymbol{n})=
	\int\limits_{\mathbb{T}^d}K(\boldsymbol{z})
	e^{-\imath\,2\pi\boldsymbol{n}\cdot\boldsymbol{z}}\mathrm{d}\boldsymbol{z}\,,
$$
and, therefore, depends only on the index $\boldsymbol{n}$.
In a further special case of a convolution integral operator having a kernel $K$ given as an inverse Fourier series,
$$
	K(\boldsymbol{z})=
	\sum_{\boldsymbol{k}\in\mathbb{Z}^d}c_K(\boldsymbol{k})
	e^{-\imath\,2\pi\boldsymbol{k}\cdot\boldsymbol{z}}\,,
$$
its symbol reduces to a Fourier coefficient
	$$
	\sigma_{\mathcal{A}}(\boldsymbol{x},\boldsymbol{n})= \sigma_{\mathcal{A}}(\boldsymbol{n})=
	c_K(-\boldsymbol{n})\,.
$$
Therefore, for real $\nu$, the Bardeen--Cooper--Schrieffer operator $\mathcal{B}$ is of the order $-\nu$
with the symbol
$$
 \sigma_{\mathcal{B}}(\boldsymbol{0})=0\quad \text{and}\quad
 \sigma_{\mathcal{B}}(\boldsymbol{n})=\dfrac{1}{|\bm n|^\nu}\,,\quad \boldsymbol{n} \ne \bm 0\,,
$$
as $\bm n=\bm 0$ is excluded from the Epstein zeta sum in \eqref{Eqn: Epstein zeta function}.

\section{Numerical approximation}
Let $\mathbb{V}$ and $\mathbb{W}$ be two Banach spaces and $\mathcal{A}\,:\, \mathbb{V}\rightarrow\mathbb{W}$
a linear and bounded operator with
$$
\|\mathcal{A}u\|_{\mathbb{W}}\leq c\,\|u\|_{\mathbb{V}}\  
\text{for all } u\in\mathbb{V}\,, c>0\,. 
$$
For a given right-hand side $f\in\mathbb{W}$, find $u\in\mathbb{V}$ which solves the following operator equation
$$
\mathcal{A}u=f\,,\ f\in\mathbb{W}\,,\ u\in\mathbb{V}\,.
$$
The variational formulation of the above equation reads as follows: find $u\in\mathbb{V}$ with
$$
\langle\mathcal{A}u,v\rangle=\langle f,v \rangle\ \text{for all } v\in\mathbb{W}'\,.
$$
Here, $\langle \cdot,\cdot \rangle\,:\,\mathbb{W}\times\mathbb{W}'\rightarrow\mathbb{C}$ denotes the duality pairing between $\mathbb{W}$ and $\mathbb{W}'$.
Let
$$
\Phi_n=(\phi_1,\dots,\phi_n)\subset\mathbb{V}\quad \text{and}\quad 
\Psi_n=(\psi_1,\dots,\psi_n)\subset\mathbb{W}'
$$
be two linear independent systems of functions and distributions and
$$
\mathbb{V}_n=\operatorname{span}\Phi_n\subset\mathbb{V}\quad \text{and}\quad 
\mathbb{W}_n'=\operatorname{span}\Psi_n\subset\mathbb{W}'
$$
the corresponding subspaces. The Galerkin-Petrov formulation of the operator equation formulation reads then: Find $u_n\in\mathbb{V}_n$ which solves the following variational formulation
$$
\langle\mathcal{A}u_n,v \rangle= \langle f,v \rangle \ \text{for all } v\in\mathbb{W}_n'\,.
$$
The formulation above is equivalent to a system of linear equations
\begin{eqnarray}
	\label{Eq: System of Linear Equations}
	Ax=b\,,\ A\in\mathbb{C}^{n\times n}\,,\quad x,b\in\mathbb{C}^n  
\end{eqnarray}
with
\begin{eqnarray}
	\label{Eq: System Enries}
	u_n=\Phi_nx\,,\quad a_{k\ell}=\langle \mathcal{A}\phi_{\ell},\psi_k \rangle\,,\quad b_k=\langle f,\psi_k \rangle\,,\quad
	k,\ell=1,\dots,n\,.
\end{eqnarray} 
In the special case $\mathbb{V}\subseteq\mathbb{W}'$, the pure Galerkin formulation is possible: Find $u_n\in\mathbb{V}_n$ which solves the following variational formulation
$$
\langle \mathcal{A}u_n,v \rangle=\langle f,v \rangle\ \text{for all } v\in\mathbb{V}_n\,.
$$
The corresponding entries of the system matrix $A$ and of the right-hand side $b$ are as follows
$$
a_{k\ell}=\langle \mathcal{A}\phi_{\ell},\phi_k \rangle\,,\quad b_k=\langle f,\phi_k \rangle\,,\quad
k,\ell=1,\dots,n\,.
$$

Consider the basis functions $\phi_{\bm \ell}\,:\,\mathbb{R}^d\rightarrow\mathbb{C}$ and their $d$-dimensional Fourier series
$$
	\phi_{\bm \ell}(\bm x)=\sum_{\bm m\in \mathbb{Z}^d}
	c_{\bm \ell}(\bm m)e^{\imath\,2\pi\, \bm m\cdot \bm x}
$$
For an integral operator $\mathcal{A}$ with the symbol $\sigma_{\mathcal{A}}$, the entries of its Galerkin matrix are in general quite complicated to compute
$$ 
    a_{\boldsymbol{k}\boldsymbol{\ell}}=\sum_{\boldsymbol{m}_1,\boldsymbol{m}_2\in\mathbb{Z}^d}
	c_{\boldsymbol{\ell}}(\boldsymbol{m}_1)\overline{c_{\boldsymbol{k}}(\boldsymbol{m}_2)}
	\int\limits_{\mathbb{T}^d}\sigma_{\mathcal{A}}(\boldsymbol{x},\boldsymbol{m}_1)
	e^{\imath\,2\pi(\boldsymbol{m}_1-\boldsymbol{m}_2)\cdot\boldsymbol{x}}\mathrm{d}\boldsymbol{x}\,.
$$
However, for a convolution operator $\mathcal{A}$ with the kernel given as an inverse Fourier series, the entries of its Galerkin matrix simplify to
$$
	a_{\boldsymbol{k}\boldsymbol{\ell}}=\sum_{\boldsymbol{m}\in\mathbb{Z}^d}
	c_{\boldsymbol{\ell}}(\boldsymbol{m})\overline{c_{\boldsymbol{k}}(\boldsymbol{m})}
	\sigma_{\mathcal{A}}(\boldsymbol{m})\,.
$$
Thus, the Galerkin method based on the Fourier series of the basis functions is very convenient for the Bardeen--Cooper--Schrieffer operator  $\mathcal{B}$.

We propose to use the multidimensional, one-periodic B-splines on $\mathbb{T}^d$ for a uniform description of families of Galerkin methods of arbitrary approximation order. 
First, we consider the one-dimensional case $d=1$ and the following discretisation of the interval $[0,1)$
$$
	[0,1)=\bigcup_{\ell=1}^n\,[x_{\ell},x_{\ell+1})\,,\quad x_{\ell}=(\ell-1)h\,,\quad h=1/n\,,\quad n\in\mathbb{N}\,,
$$
where $n$ and, therefore, $h$ will be the same for all space dimensions. A possible basis
in the space of piecewise polynomials on $[0,1)$ with respect to the above decomposition
and periodically extended to the whole $\mathbb{R}$ is denoted by 
$$
  \Phi_n^{(\mu)}=\Big(\phi_1^{(\mu)},\dots,\phi_n^{(\mu)}\Big)\,,
$$
where $\mu$ denotes the degree of the splines. The functions $\phi_{\ell}^{(\mu)}$ will be defined recursively
$$
\phi_1^{(0)}(x)=
\begin{cases}
	1 & \text{for}\quad -h/2+m\leq x < h/2+m\,,\quad m\in\mathbb{Z}\,,\\
	0 & \text{elsewhere}. \\
\end{cases}\,
$$
For $\mu\geq 1$ and for $n>\mu+1$, we define the B-splines of the order $\mu$ by the use of the recursion
$$
\phi_1^{(\mu)}(x)=\frac{1}{h}\,\int\limits_{-1/2}^{1/2}
\phi_1^{(\mu-1)}(y)\phi_1^{(0)}(x-y)\,\mathrm{d}y\,,\quad \mu=1,2,\dots\ ,
$$
and
$$
\phi_{\ell}^{(\mu)}(x)=\phi_1^{(\mu)}(x-x_{\ell})\,,\quad \ell=2,\dots,n\,,
$$
as well as by the periodical extension
$$
\phi_{\ell}^{(\mu)}(x+m)=\phi_{\ell}^{(\mu)}(x)\,,\quad m\in\mathbb{Z}\,.
$$
The finite dimensional space of all periodic piecewise polynomials on $\mathbb{R}$
will be denoted by
$$
{H}_n=\operatorname{span}\,\Phi_n^{(\mu)}=
\big\{u_n=\Phi_n^{(\mu)}c\,,\ c\in\mathbb{C}^n\big\}\,.
$$
The basic properties of the B-splines defined above are summarised in the following theorem.
\begin{theorem}
	\label{Thm: B-splines Properties}
The main properties of the 1-periodic B-Splines are:
		\begin{enumerate}
		\item Decomposition of unity: 
		$$
		\Phi_n^{(\mu)}e=\sum_{\ell=1}^n\phi_{\ell}^{(\mu)}(x)=1\,,\quad \text{for all}\quad
		x\in\mathbb{R},\,\quad n>\mu+1\,.
		$$
		\item The first B-splines are even functions:
		$$
		\phi_1^{(\mu)}(x)=\phi_1^{(\mu)}(-x)\,,\quad \text{for all}\quad 
		x\in\mathbb{R}\,,\quad \mu=0,1,\dots\,.
		$$
		\item The Fourier coefficients:
		$$
		\phi_{\ell}^{(\mu)}(x)=\sum_{m\in\mathbb{Z}}c_{\ell}^{(\mu)}(m)\,e^{\imath\,2\pi\, m x}\,,
		$$
		where
		$$
		c_1^{(\mu)}(m)=h\,\operatorname{sinc}^{\mu+1}(\pi h\,m)\,,
		$$
		with
		$$
		\operatorname{sinc}(x)=
		\begin{cases}
			1 & \text{for}\quad x=0\,, \\[1ex]
			\dfrac{\sin(x)}{x} & \text{for}\quad x\neq 0\,,\quad x\in\mathbb{R}, \\
		\end{cases}
		$$
		and
		$$
		c_{\ell}^{(\mu)}(m)=c_1^{(\mu)}(m)e^{-\imath\,2\pi h\,m(\ell-1)}\,,\quad \ell=2,\dots,n\,.
		$$
		\item Sobolev spaces:
		$$
		\phi_{\ell}^{(\mu)}\in{H}^a\quad  \text{for all}\quad a<\mu+1/2\,.
		$$
		\item Approximation property:
		$$
		\inf_{u_n\in{H}_n}\|u-u_n\|_{{H}^a}\leq c_1\,h^{s-a}\|u\|_{{H}^s}\quad \text{for all}\quad u\in{H}^s\,,\quad c_1>0\,,
		$$
		where
		$$
		-\infty < a \leq s\leq \mu+1\,,\quad a < \mu+1/2\,.
		$$
		\item Inverse inequality: 
		$$
		c_2\,h^{s-a}\|u_n\|_{{H}^s}\leq \|u_n\|_{{H}^a}\leq \|u_n\|_{{H}^s}\quad  \text{for all}\quad  u_n\in{H}_n\,,  
		$$
		where
		$$
		-\infty < a \leq s < \mu+1/2\,.
		$$
	\end{enumerate}
\end{theorem}
In multidimensional case $d>1$, we simply define 
$$
\phi_{\boldsymbol{\ell}}^{(\mu)}(\boldsymbol{x})=\prod_{j=1}^d\phi_{\ell_j}^{(\mu)}(x_j)\,,\quad 
\bm \ell \in \{1,\dots,n\}^d\,,\quad \boldsymbol{x}\in\mathbb{R}^d\,.
$$
Due to this definition, the Fourier coefficients of the basis functions $\phi_{\boldsymbol{\ell}}^{(\mu)}$ are given by
$$
c_{\boldsymbol{\ell}}^{(\mu)}(\boldsymbol{m})=\prod_{j=1}^dc_{\ell_j}^{(\mu)}(m_j)\,,\quad  
\bm \ell \in \{1,\dots,n\}^d\,,\quad \boldsymbol{m}\in\mathbb{Z}^d\,.
$$
Furthermore, with the third property of the one-dimensional Fourier coefficients, we get
$$
c_{\boldsymbol{\ell}}^{(\mu)}(\boldsymbol{m})=
c_{\boldsymbol{e}}^{(\mu)}(\boldsymbol{m})
e^{-\imath\,2\pi h\,\boldsymbol{m}\cdot(\boldsymbol{\ell}-\boldsymbol{e})}\,,
$$
where $\boldsymbol{e}=(1,\dots,1)^{T}\in\mathbb{Z}^d$.

For a convolution operator $\mathcal{A}$, the use of the B-Splines as a basis and test functions leads to the Galerkin matrix entries of a very special form
$$
	a_{\boldsymbol{k}\boldsymbol{\ell}}=a(\boldsymbol{k}-\boldsymbol{\ell})=
	\sum_{\boldsymbol{m}\in\mathbb{Z}^d}
	\left|c_{\boldsymbol{e}}^{(\mu)}(\boldsymbol{m})\right|^2
	\sigma_{\mathcal{A}}(\boldsymbol{m})
	e^{\imath\,2\pi h\,\boldsymbol{m}\cdot(\boldsymbol{k}-\boldsymbol{\ell)}}\,.
$$
	The function $a\,:\, \mathbb{Z}^d\rightarrow \mathbb{C}$ is $n-$periodic
$$
	a(\boldsymbol{k}+n\boldsymbol{m})=a(\boldsymbol{k})\quad  
	\text{for all}\quad \boldsymbol{k},\boldsymbol{m}\in\mathbb{Z}^d\,.
$$
The matrices of this kind are called circulant matrices for $d=1$ and block-circulant
matrices for $d>1$. These matrices allow for extremely efficient, almost optimal performance in all algebraic operations, including multiplication with a vector, linear systems, and eigenvalue problems \cite{RJASANOW17}.

The BCS equation in the scalar and one-dimensional case reads
\[
f(x)=C_1\int\limits_0^1g(y)\mathrm{d}y+
C_2\int\limits_0^1Z_{\mathbb{Z},\nu}
(x-y)\,
g(y)\mathrm{d}y=C_1\,\varrho[g]+C_2\,\mathcal{B}[g](x)\,,
\]
where $\varrho$ is a simple linear functional and $\mathcal{B}$ is a pseudo-differential operator.
The function $g$ is connected with the unknown function $f$ via the following non-linear algebraic relation,
$$
  g(x)=g[f](x)=\dfrac{f(x)}{\sqrt{\cos^2(2\pi\,x)+|f(x)|^2}}\,.
$$
Thus, we have to solve a system of two equations for unknown functions $f,g$
\begin{align*}
	f&=C_1\,\varrho[g]+C_2\,\mathcal{B}[g]\,, \\
	g&=g[f]\,.
\end{align*}
Let ${H}_n=\operatorname{span}\,\Phi_n^{(\mu)}$ be a $n-$dimensional B-spline subspace. Then, the Galerkin formulation of the above system reads: find $f_n\in{H}_n$ and $g_n\in{H}_n$ such that
\begin{align*}
	\langle f_n,v \rangle&=C_1\,\varrho[g_n]\langle 1,v\rangle+C_2\,\langle\mathcal{B}[g_n],v\rangle\,, \\[1ex]
	\langle g_n,v \rangle &=\langle g[f_n],v \rangle \quad \text{for all}\quad v\in {H}_n\,.
\end{align*}
With
$$
f_n=\Phi_n^{(\mu)}\underline{f}\,,\quad g_n=\Phi_n^{(\mu)}\underline{g}\,,\quad 
\underline{f},\underline{g}\in\mathbb{C}^n\,,
$$
we formulate the following algebraic system
\begin{align*}
	M\underline{f}&=A\underline{g}\,, \\
	M\underline{g}&=G(\underline{f})\underline{f}\,.
\end{align*}
The circulant mass matrix $M=M^{(\mu)}\in\mathbb{R}^{n\times n}$ is given as
$$
M_{k\ell}^{(\mu)}=\langle \phi_{\ell}^{(\mu)},\phi_k^{(\mu)} \rangle=
\int\limits_0^1
\phi_{\ell}^{(\mu)}(x)\phi_k^{(\mu)}(x)\mathrm{d}x\,,\quad 
k,\ell=1,\dots,n
$$
is, additionally, symmetric, positive definite and of the periodic band structure.
For $\mu=0$, we get
$$
M^{(0)}=
h\,I=h\,\operatorname{circ}(1,0,\dots,0)\in\mathbb{R}^{n\times n}\,,
$$ 
for $\mu=1$
$$
M^{(1)}=\dfrac{1}{6}\,h\,
\operatorname{circ}(4,1,0,\dots,0,1)\in\mathbb{R}^{n\times n}\,,
$$
and for $\mu=2$
$$
M^{(2)}=\dfrac{1}{120}\,h\,
\operatorname{circ}(66,26,1,0,\dots,0,1,26)\in\mathbb{R}^{n\times n}\,.
$$
The matrix $A=A^{(\mu)}\in\mathbb{R}^{n\times n}$
is of the form 
$$
A^{(\mu)}=C_1h^2E+C_2B^{(\mu)}
$$
with
$$
E_{k\ell}=1\quad \text{and}\quad 
B_{k\ell}^{(\mu)}=
\langle \mathcal{B}[\phi_{\ell}^{(\mu)}],\phi_k^{(\mu)}\rangle=
\int\limits_0^1\int\limits_0^1Z_{\mathbb{Z},\nu}(x-y) \,
\phi_{\ell}^{(\mu)}(y)\phi_k^{(\mu)}(x)
\mathrm{d}y\,\mathrm{d}x\,.
$$
The matrix $A=A^{(\mu)}$ is circulant, symmetric and positive definite as well.
Its entries can be computed efficiently by the use the symbol of the operator $\mathcal{B}$ and of the Fourier coefficients of the basis functions.
If we  denote by
$$
M(x)=M^{(\mu)}(x)\in\mathbb{R}^{n\times n}\quad \text{with}\quad
M_{k\ell}^{(\mu)}(x)=\phi_{\ell}^{(\mu)}(x)\phi_k^{(\mu)}(x)
$$
the functional mass matrix, which is obviously symmetric and positive definite, then the matrix $G(\underline{f})=G^{(\mu)}(\underline{f})$ defined by 
$$
G^{(\mu)}(\underline{f})_{k\ell}=
\int\limits_0^1
\dfrac{\phi_{\ell}^{(\mu)}(x)\phi_k^{(\mu)}(x)}{\sqrt{\xi^2(x)+\underline{f}^\ast M^{(\mu)}(x)\underline{f}}}\,
\mathrm{d}x
$$
is symmetric, positive definite and of the same periodic band structure as the mass matrix.
During the iterative solution of the above algebraic system, this matrix will need to be recomputed at every iteration step. However, as this matrix is very sparse and circulant, the computational cost is minimal.
\section{Numerical example}

Finally, we present the numerical solution to the two-dimensional long-range BCS equation on the square lattice $\Lambda=\mathbb Z^2$ for $C_1=0.75,\ C_2=0.7$, and $\nu=2.01$. For this choice of parameters, the gap matrix $F$ admits the following structure,
$$
 F(\bm x)=\begin{pmatrix} 0 & f(\bm x) \\ -f(-\bm x) & 0 \end{pmatrix}\,,\quad
 \operatorname{Im}f(\bm x)=0\,.
$$ 
Hence, the problem is reduced to a real-valued scalar function $f$.

\begin{figure}
	\begin{center}
		\includegraphics[width=7cm]{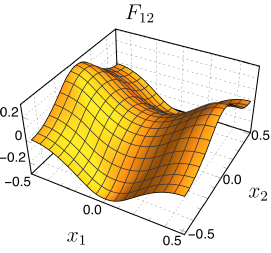}
	\end{center}
	\caption{Nodal d-wave solution of the gap equation for $C_1=0.75,\ C_2=0.7$, and $\nu=2.01$}.
    \label{Fig:d-wave}
\end{figure}

The solution shown in Figure \ref{Fig:d-wave} was computed using periodic cubic B-splines on a two-dimensional $1024 \times 1024$ grid. According to the standard symmetry classification of gap solutions \cite{sigrist1991phenomenological}, it is identified as a $d$-wave solution,
\[
f(\bm x)\sim \cos(2\pi x_1)-\cos(2\pi x_2),
\]
which frequently appear in the study of unconventional superconductors. In contrast to a standard $s$-wave solution, where $f$ has the symmetries of a constant function, a $d$-wave solution changes sign when exchanging $x_1$ and $x_2$. In particular, the function $f$ exhibits zeros, which leads to important physical and numerical consequences. At points $\bm x$, where both $f(\bm x)$ and the dispersion relation $\xi(\bm x)$ vanish, the nonlinear function $G[F]$ in \eqref{Eqn: Direct F->G} exhibits a discontinuity. We speak of nodal superconductivity, which occurs here, i.e., at $\bm x = (1/4,1/4)^\top$. Correctly resolving these singular points during the convolution with the singular Epstein zeta function is numerically challenging. Using a discretisation of the reciprocal unit cell with periodic B-splines, along with an analytic evaluation of the convolution, allows our algorithm to resolve these discontinuities, thereby enabling a precise study of the interplay of long-range interactions and nodal superconductivity.

\section*{Acknowledgement}
	We would like to express our sincere gratitude to Professor Roland Duduchava for his friendship and scientific collaboration over many years, his co-supervision of doctoral students, and his numerous invited lectures in Saarbrücken. 
    
    This work was supported by the Klaus-Tschira Stiftung under Grant No. 00.025.2025. 
    
    The authors received scientific support and HPC resources provided by the Erlangen National High Performance Computing Center (NHR@FAU) of the Friedrich-Alexander-Universität Erlangen-Nürnberg (FAU) under the NHR project n101af.


\begin{thebibliography}{99}

\bibitem{BCS}
J.~Bardeen, L.~N.~Cooper, and J.~R.~Schrieffer.
\newblock {Microscopic Theory of Superconductivity}.
\newblock {\em Phys. Rev.}, 106:162--164, 1957.

\bibitem{buchheit2024epstein}
A.~A.~Buchheit, J.~Busse, and R.~Gutendorf. 
Computation and properties of the Epstein zeta function with high-performance implementation in EpsteinLib.
In: {\em arXiv}, 2412.16317, 2024.

\bibitem{buchheit2023exact}
A.~A.~Buchheit, T.~Ke\ss{}ler, P.~K.~Schuhmacher, and B.~Fauseweh.
\newblock Exact continuum representation of long-range interacting systems and
  emerging exotic phases in unconventional superconductors.
\newblock {\em Phys. Rev. Res.}, 5:043065, 2023.

\bibitem{covaci2010efficient}
L.~Covaci, F.~M.~Peeters, and M.~Berciu.
\newblock Efficient numerical approach to inhomogeneous superconductivity: The
  Chebyshev--Bogoliubov--de Gennes method.
\newblock {\em Phys. Rev. Lett.}, 105(16):167006, 2010.

\bibitem{epstein1903theorieI}
P.~Epstein.
\newblock {Zur Theorie allgemeiner Zetafunktionen}.
\newblock {\em Math. Ann.}, 56:615–644, 1903.

\bibitem{epstein1903theorieII}
P.~Epstein.
\newblock {Zur Theorie allgemeiner Zetafunktionen. II}.
\newblock {\em Math. Ann.}, 63:205--216, 1906.

\bibitem{fanfarillo2025superconductivity}
L.~Fanfarillo, Y.~Cao, C.~Setty, S.~Caprara, and P.~J.~Hirschfeld.
\newblock Superconductivity with repulsion: A variational approach.
\newblock {\em Phys. Rev. B}, 112(13):134519, 2025.

\bibitem{haink2025non}
D.~Haink, A.~A.~Buchheit, and B.~Fauseweh. 
Non-local edge mode hybridization in the long-range interacting Kitaev chain. In: {\em arXiv}, 2509.26447, 2025.

\bibitem{hainzl2023boundary}
C.~Hainzl, B.~Roos, and R.~Seiringer.
\newblock Boundary superconductivity in the BCS model.
\newblock {\em J. Spectr. Theory}, 12(4):1507--1540, 2023.

\bibitem{hainzl2016bardeen}
C.~Hainzl and R.~Seiringer.
\newblock The Bardeen--Cooper--Cchrieffer functional of superconductivity and
  its mathematical properties.
\newblock {\em J. Math. Phys.}, 57(2):021101, 2016.

\bibitem{RJASANOW17}
S.~Rjasanow.
\newblock Effective algorithms with circulant-block matrices.
\newblock {\em Linear Algebra Appl.}, 202:55--69, 1994.

\bibitem{Vainikko}
J.~Saranen and G.~Vainikko.
\newblock {\em Periodic integral and pseudodifferential equations with
  numerical approximation}.
\newblock Springer-Verlag, Berlin, 2002.

\bibitem{sigrist1991phenomenological}
M.~Sigrist and K.~Ueda.
\newblock Phenomenological theory of unconventional superconductivity.
\newblock {\em Rev. Mod. Phys.}, 63(2):239, 1991.

\bibitem{stewart2017unconventional}
G.~R.~Stewart.
\newblock Unconventional superconductivity.
\newblock {\em Adv. Phys.}, 66(2):75--196, 2017.

\end{thebibliography}

\end{document}